# Space-Efficient Bounded Model Checking

Jacob Katz[1], Ziyad Hanna[1], Nachum Dershowitz[2]

[1] Intel Corporation, Haifa, {jacob.katz, ziyad.hanna}@intel.com

[2] School of Computer Science, Tel Aviv University, Tel Aviv, Israel,
nachum.dershowitz@cs.tau.ac.il

**Abstract**

*Current algorithms for bounded model checking use SAT methods for checking satisfiability of Boolean formulae. These methods suffer from the potential memory explosion problem. Methods based on the validity of Quantified Boolean Formulae (QBF) allow an exponentially more succinct representation of formulae to be checked, because no "unrolling" of the transition relation is required. These methods have not been widely used, because of the lack of an efficient decision procedure for QBF. We evaluate the usage of QBF in bounded model checking (BMC), using general-purpose SAT and QBF solvers. We develop a special-purpose decision procedure for QBF used in BMC, and compare our technique with the methods using general-purpose SAT and QBF solvers on real-life industrial benchmarks.*

## 1 Introduction[*]

Model checking is a technique for the verification of the correctness of a finite-state system with respect to a desired behavior. Symbolic model checking uses image computation to verify properties. Symbolic model checking methods include, among others, BDD-based techniques, SAT-based methods for image computation that use an explicit quantifier elimination, and SAT-based reachability analysis based on "all-solutions" SAT solvers. All these methods suffer from the memory explosion problem on modern test cases.

Bounded Model Checking (BMC) with a specific bound $k$ represents the paths of length $k$ in the system by "unrolling" the transition relation $k$ times, and examines whether the set of states falsifying the property is reached by these paths. To implement a complete model checking procedure the bound should be increased iteratively up to the length of the longest simple path in the system, causing the number of copies of the transition relation within the formulae being checked for validity to increase from iteration to iteration up to an exponential number of times,

---

[*] Due to space constraints references have been omitted in this text.

leading, again, to a memory explosion for large systems and large bounds.

Induction based methods provide another technique for estimating whether a bound is sufficient to ensure a full proof, but there are still many cases where the induction depth is exponential in the size of the model.

Finally, the methods based on Craig interpolation as an over-approximation technique for image computation aimed at reducing the number of iterations for a complete model checking procedure. The interpolants are obtained as a by-product of the SAT solver used to check BMC problems. This technique, like other techniques based on image computation, also suffers from a potential memory blow-up.

In this paper we present a short abstract of our research on the usage of Quantified Boolean Formulae (QBF) for BMC, in which the unrolling of the transition relation is not performed and, thus, the memory explosion problem is avoided. We evaluate available general-purpose QBF solvers, and develop a special-purpose decision procedure for QBF used in BMC. We also compare our technique with the classical SAT-based BMC methods.

## 2 Formulations of bounded reachability checking problem

Given a system M=(S, I, TR), where S is the set of states, I is the characteristic function of the set of the initial states, and TR is the transition relation, the problem of reachability of the final states given by a characteristic function F in exactly $k$ steps can be expressed in a number of ways.

As in classical BMC, the fact that the state $Z_k$ is reachable from the state $Z_0$ in exactly $k$ steps may be formulated by "unrolling" the transition relation $k$ times:

(1) $$R_k(Z_0, Z_k) = \exists Z_1,...,Z_{k-1} : I(Z_0) \wedge F(Z_k) \wedge \bigwedge_{i=0}^{k-1} TR(Z_i, Z_{i+1})$$

The validity of this formula may be proven or disproven by performing the SAT decision procedure on its propositional part. Noticeably, the number of copies of the transition relation in this formula is as the number of steps being checked. To partially overcome the potential memory explosion, a QBF formulation of bounded reachability problem can be used:

$$(2) \quad R_k(Z_0, Z_k) = \exists Z_1, ..., Z_{k-1} : I(Z_0) \wedge F(Z_k) \wedge \\ \forall U, V : \left( \bigvee_{i=0}^{k-1} (U \leftrightarrow Z_i) \wedge (V \leftrightarrow Z_{i+1}) \right) \rightarrow TR(U, V)$$

Note that (2) contains only one copy of the transition relation. Increasing the bound, thus, would mean an addition of a new intermediate state and a term of the form $(U \leftrightarrow Z_i) \wedge (V \leftrightarrow Z_{i+1})$. Hence, the formula increase from iteration to iteration does not depend on the size of the transition relation, which is usually the biggest formula in the specification of the model.

The solution of (2) with a QBF solver usually requires a transformation of the propositional part of the formula into a CNF, which introduces artificial variables, resulting with a QBF having $\exists \forall \exists$ pattern of the quantifier prefix. The number of the universally quantified variables does not change in the QBF from iteration to iteration.

This approach to reachability checking partially solves the issue of formula growth, reducing the growth of the formula from iteration to iteration, but still requires an exponential number of iterations to fully verify the reachability.

To reduce the number of iterations, it is possible to apply the "iterative squaring" technique, similar to the one used in BDD-based model checking. In this technique, each successive iteration checks the reachability of a final state in twice as many steps as the previous iteration. Given a formula $R_{k/2}(X,Y)$ for checking reachability in $k/2$ steps, the following formula checks the reachability in $k$ steps:

$$(3) \quad R_k(Z_0, Z_k) = \exists Z : I(Z_0) \wedge F(Z_k) \wedge \forall U, V : \\ \left[ (U \leftrightarrow Z_0) \wedge (V \leftrightarrow Z) \vee (U \leftrightarrow Z) \wedge (V \leftrightarrow Z_k) \right] \rightarrow R_{k/2}(U, V)$$

The transition relation appears in (3) only once, as in the previously described technique. However, the number of universally quantified variables and the number of quantifier alternations grows from iteration to iteration.

This technique allows reducing the number of iterations to be as the number of the state encoding variables in the model. Note that not all bounds are checked by this technique, but only the bounds that are a power of 2. It is possible, however, to overcome this problem by adding a self-loop in each state of the model, which would not change the reachability between states, but rather make (3) check reachability in *k or fewer* steps, instead of *exactly k* steps.

## 3 BMC using QBF

We have used a bounded model checker to generate the three kinds of formulae mentioned in the previous section. We have evaluated a few available state-of-the-art DPLL-based SAT and QBF solvers, to check the feasibility of the QBF formulations of the reachability checking problem on a set of thirteen proprietary Intel® model checking test cases of different sizes. It appeared that the general-purpose QBF solvers were unable to solve practically any of the formulae of the forms (2) and (3), while many of the corresponding propositional formulae of the form (1) were solved by the SAT solvers, the majority of them in a matter of seconds.

Motivated by the inefficiency of the general-purpose QBF solvers demonstrated on formulae of the form (2), we develop a special-purpose DPLL-based decision procedure, called jSAT, for formulae of this specific structure. As in (2), jSAT holds in memory the encoding variables representing the states $Z_0, Z_1, ..., Z_k$, U and V, but only holds the following propositional formula:

$$(4) \quad I(Z_0) \wedge TR(U, V) \wedge F(Z_k)$$

The states $Z_i$ represent a path; the states U and V represent two neighboring states in that path. Instead of explicitly holding the fact that U and V represent a pair of neighboring states as done in (2) with assistance of the terms of the form $(U \leftrightarrow Z_i) \wedge (V \leftrightarrow Z_{i+1})$, our algorithm implicitly assumes this information. The idea of the algorithm is to iteratively associate U and V with a pair of successive states, called the current state and the next state, until all states are decided.

Intuitively, jSAT algorithm can be seen as a depth-first search in the state graph of the system from the initial states to the final ones. The algorithm starts by associating U with $Z_0$ and V with $Z_1$; thus the formula (4) becomes semantically equivalent to:

$$(5) \quad I(Z_0) \wedge TR(Z_0, Z_1) \wedge F(Z_k)$$

The states $Z_0$ and $Z_1$ are then decided, if possible, so that $Z_0$ is an initial state and $Z_1$ is its successor. As soon as they are decided, the algorithm makes $Z_1$ to be the current state and $Z_2$ to be the next one. The algorithm proceeds so on, until all states are successfully decided, or until it discovers that such a decision is impossible.

The first implementation of our algorithm succeeded to solve 143 out of 234 instances of the form (2) in our test base, compared to 184 corresponding SAT instances solved by the solver on which we based our implementation, and 3 instances solved by the general purpose QBF solvers, within 300 seconds time limit and 1GB memory limit.